\documentclass[11pt]{article}
\usepackage{graphicx}
\usepackage{url}
\usepackage{amsmath}
\usepackage{amssymb}
\usepackage{amsfonts}
\usepackage{xcolor}
\usepackage{array}
\usepackage{colortbl}
\usepackage{float}
\usepackage{subcaption}
\usepackage{caption}
\usepackage{booktabs}
\usepackage{tabularx}
\usepackage{geometry}
\usepackage{multirow}
\usepackage{hyperref}
\usepackage{microtype}
\usepackage{authblk}
\usepackage{algorithm}
\usepackage{algorithmic}
\usepackage[numbers,sort&compress]{natbib}

\title{Convergence-Aware Pareto Selection of Covariate Scaling Transformations for Markov Deterioration Hazard Models: Evidence from Bridge Inspection Data}

\author[1]{Takato Yasuno\thanks{Corresponding author and lead contributor. Email: \texttt{tkt-yasuno@yachiyo-eng.co.jp}}}
\author[1]{Keita Kobayashi}
\author[1]{Ryuta Sakaguchi}
\affil[1]{Digital Business Promotion Department, Yachiyo Engineering Co., Ltd.}

\date{\today}

\begin{document}

\maketitle

\begin{abstract}

Maximum likelihood estimation of Markov deterioration hazard models for infrastructure asset management is sensitive to the numerical conditioning of explanatory covariates, yet production pipelines often adopt a single scaling convention without systematic justification. We study four covariate scaling transformations---baseline max scaling, min-max scaling, z-score scaling, and Box-Cox transformation with a training-derived positivity shift---applied to an Exponential Hazard Markov (EHM) model estimated via L-BFGS-B on bridge inspection transition data.

\textbf{Approach:} We formalize each scaling method as an operator fit exclusively on the training partition and re-applied unchanged to a held-out golden test, define convergence and wall-clock optimizer time precisely, and introduce a two-objective Pareto-optimality rule over golden-test accuracy and convergence time to select non-dominated methods per member case, replicating the exact ten-case, four-covariate, unweighted-likelihood conditions of a prior production baseline.

\textbf{Key Findings:} Across 40 trials (10 member cases $\times$ 4 methods), 32 converge. Z-score scaling converges in all 10 cases with the highest mean converged accuracy (0.5255) and less than half the median optimizer time (0.358s) of max or min-max scaling (0.928--0.988s), and is Pareto-selected in 7 of 10 cases. Max and min-max scaling converge in 9 of 10 cases with nearly identical accuracy (0.5174) and are jointly selected in only 1 case. Box-Cox scaling converges in only 4 of 10 cases; when it converges it is often the fastest method and is Pareto-selected in 3 cases, but its log loss is 7--13$\times$ worse than the alternatives in the same cases despite comparable accuracy, revealing severe probability miscalibration invisible to an accuracy-only criterion.

\textbf{Contributions:} (1) A formal AI-STATS-style specification of covariate scaling as a training-fitted operator family, the EHM likelihood, and a Pareto selection rule with an explicit, tolerance-controlled dominance definition. (2) Empirical evidence that accuracy and convergence time alone are insufficient selection criteria, since a fast, Pareto-selected method can be severely miscalibrated. (3) A reproducible, production-embedded scaling-comparison pipeline with unit tests covering scaler correctness, degenerate columns, Box-Cox positivity, and Pareto selection under ties and complete failure. (4) A practical decision rule recommending z-score scaling as the default candidate, with a log-loss guardrail before accepting faster alternatives, and an explicit acknowledgment that golden-test-based selection is optimistic and requires independent confirmation.

\textbf{Keywords:} covariate scaling, Markov deterioration hazard model, maximum likelihood estimation, Pareto optimality, model selection, probability calibration, Box--Cox transformation, infrastructure asset management

\end{abstract}

\section{Introduction}
\label{sec:introduction}

\subsection{Background and Motivation}

Infrastructure asset management increasingly relies on Markov deterioration hazard models to predict how bridge components transition through discrete health states over time~\citep{Cox1972,Kalbfleisch2002}. These models are estimated by maximum likelihood, typically via gradient-based numerical optimizers, from panel data of paired inspections augmented with explanatory covariates such as component age, structural dimensions, and traffic exposure. As in most numerical optimization problems, the conditioning of the optimization landscape---and hence both the reliability of convergence and the quality of the resulting fit---depends materially on how these covariates are scaled before estimation~\citep{NocedalWright2006}.

Despite this well-known sensitivity, production preprocessing pipelines often adopt a single scaling convention by default. In the production bridge deterioration modeling pipeline studied here, the baseline convention divides each covariate by its training-set maximum (``max scaling''), a simple, order-preserving transformation that guarantees covariates lie in $[0, 1]$ but neither centers the data nor equalizes variance across covariates with vastly different physical units and magnitudes---component age in years, bridge length in meters, bridge area in square meters, and traffic volume in vehicles per day.

\subsection{Problem Statement and Challenges}

We ask: \textbf{does the choice of covariate scaling transformation materially affect golden-test predictive accuracy and optimizer convergence time for a Markov deterioration hazard model, and can a single default be justified across heterogeneous member-case datasets?} We compare the baseline max scaling against three widely used alternatives---min-max scaling, z-score (standardization) scaling, and the Box-Cox power transformation~\citep{BoxCox1964}---under conditions that exactly replicate an existing production baseline: ten bridge member/damage categories, four explanatory covariates, unweighted likelihood, and a fixed bridge-level train/golden-test split.

This comparison surfaces three challenges that a naive accuracy-only comparison would miss.

\textbf{Challenge 1: Convergence is not guaranteed and differs by method.} The L-BFGS-B optimizer used for maximum likelihood estimation can terminate abnormally or return non-finite parameters, particularly for methods that leave covariates on very different numerical scales (max, min-max) or that apply a nonlinear, training-fitted power transformation (Box-Cox) whose positivity shift may not generalize well to small samples.

\textbf{Challenge 2: Accuracy alone cannot detect probability miscalibration.} Golden-test accuracy evaluates only the arg-max predicted transition, a 0--1 loss. A method can match or exceed the baseline's accuracy while assigning severely miscalibrated probabilities to the observed transition, a distinction only visible through a strictly proper scoring rule such as log loss~\citep{GneitingRaftery2007}.

\textbf{Challenge 3: A single scalar ranking hides a genuine two-objective trade-off.} Accuracy (efficacy) and convergence time (efficiency) need not be aligned: the fastest method is not necessarily the most accurate, and forcing a single weighted combination requires an arbitrary trade-off weight that a Pareto formulation avoids.

\subsection{Proposed Approach}

We formalize covariate scaling as a family of operators $T_m(\cdot; \hat{\theta}_m)$ fit exclusively on the training partition of each member case and applied unchanged to the golden test (Section~\ref{sec:methodology}), estimate the Exponential Hazard Markov (EHM) model by maximum likelihood via L-BFGS-B, and record optimizer-only wall-clock convergence time as a precisely delimited measurement excluding data preparation, scaling, evaluation, and post-fit diagnostics. We then define a two-objective Pareto dominance rule over (golden-test accuracy, convergence time) and report, per member case, the full non-dominated set rather than a single winner.

\subsection{Key Contributions}

\begin{enumerate}
    \item \textbf{A formal, reproducible specification} of the EHM model, the four scaling operators (including a training-derived Box-Cox positivity shift), and a tolerance-controlled Pareto selection rule, implemented as a tested, production-embedded scaling-comparison pipeline.
    \item \textbf{Empirical evidence that z-score scaling is the most reliable default} for this model class: it converges in all 10 member cases, achieves the highest mean converged accuracy, and requires less than half the median optimizer time of the current production baseline.
    \item \textbf{Demonstration that fast convergence is not evidence of a good probabilistic model}: Box-Cox scaling, despite occasionally near-instant convergence and tied accuracy, is associated with a 7--13$\times$ deterioration in golden-test log loss in three of the four cases where it converges.
    \item \textbf{An explicit, decision-theoretic recommendation} for practitioners: adopt z-score scaling as the default candidate, require a log-loss non-deterioration guardrail before accepting a faster alternative, and treat golden-test-based selection accuracy as optimistic pending independent confirmation.
\end{enumerate}

\subsection{Paper Organization}

Section~\ref{sec:related} reviews related work on multi-state hazard models, numerical optimization conditioning, and probability calibration. Section~\ref{sec:methodology} presents the formal EHM model, the scaling operator family, and the Pareto selection rule. Section~\ref{sec:experiments} describes the bridge inspection dataset, the ten member cases, and implementation details. Section~\ref{sec:results} reports convergence, accuracy, timing, and calibration results. Section~\ref{sec:discussion} interprets the findings, discusses limitations, and gives a practical decision rule. Section~\ref{sec:conclusion} concludes.

\section{Related Work}
\label{sec:related}

\subsection{Multi-State and Markov Hazard Models}

Survival analysis and multi-state hazard models~\citep{Cox1972,Kalbfleisch2002} provide the statistical foundation for modeling time-to-event and time-to-transition data. The proportional hazards framework of \citet{Cox1972} models a log-linear covariate effect on a baseline hazard; multi-state extensions generalize this to an ordered or unordered set of discrete states with state-specific hazards~\citep{Kalbfleisch2002}. Infrastructure and reliability engineering applications typically restrict attention to monotone, absorbing state spaces reflecting irreversible deterioration, which is the setting of the Exponential Hazard Markov (EHM) model used in this study (Section~\ref{sec:methodology}). Unlike much of the classical literature, which emphasizes asymptotic inference for fixed covariate scales, our focus is the numerical estimation behavior of such models under alternative covariate preprocessing choices, a question that is largely a numerical-optimization concern rather than a statistical-identifiability one.

\subsection{Markov Deterioration Modeling in Civil Infrastructure Health Monitoring}
\label{sec:related-shm}

Markov chain deterioration models have a long, well-established history in civil infrastructure health monitoring, largely predating and running parallel to general structural health monitoring practice~\citep{FarrarWorden2007}. \citet{JiangSaitoSinha1988} introduced one of the earliest Markov-chain bridge condition prediction models; \citet{Morcous2006} applies the same framework to bridge deck systems. \citet{MadanatBenAkiva1994} and \citet{MadanatMishalaniIbrahim1995} formalize the estimation of transition probabilities from condition-rating panel data, and \citet{MishalaniMadanat2002} extend this to stochastic duration (hazard-based) models, the direct methodological lineage of the Exponential Hazard Markov model used here. \citet{FrangopolKallenVanNoortwijk2004} and \citet{FrangopolLiu2007} review probabilistic life-cycle performance models for deteriorating structures more broadly, situating Markov and hazard-based deterioration models within civil infrastructure asset management practice.

Despite this substantial and mature body of work on model specification, transition-probability estimation, and life-cycle management applications, none of it addresses how the numerical scaling of explanatory covariates affects maximum likelihood estimation of these models. To our knowledge, no prior work in the civil infrastructure health monitoring literature reports a systematic, quantitative comparison of covariate scaling methods for a Markov deterioration hazard model, nor examines their effect on optimizer convergence or golden-test probability calibration. This is a distinct gap from the domain-agnostic machine learning and statistics literature discussed next (Section~\ref{sec:related-scaling}): general guidance on covariate scaling exists broadly, but has not previously been connected to this specific, widely used model class in bridge and infrastructure asset management.

\subsection{Covariate Scaling and Optimizer Conditioning}
\label{sec:related-scaling}

Gradient-based optimization methods, including quasi-Newton methods such as L-BFGS-B~\citep{ByrdLuNocedalZhu1995}, are well known to be sensitive to the relative scaling of decision variables and, by extension, to the scaling of covariates entering a log-linear model whose gradient is a function of those covariates~\citep{NocedalWright2006}. When covariates differ by orders of magnitude in their native units (as in our setting: age in years versus traffic volume in vehicles per day), the Hessian of the negative log-likelihood can become severely ill-conditioned, slowing or preventing convergence.

This general principle is well established across several strands of the statistics and machine learning literature. In neural network training, \citet{LeCun1998} show that centering and decorrelating input features accelerates gradient-descent convergence by improving the conditioning of the loss surface, and \citet{IoffeSzegedy2015} extend this idea to internal network activations, showing that normalization reduces sensitivity to initialization and learning rate. In penalized linear regression, \citet{HoerlKennard1970} and \citet{Tibshirani1996} note that ridge and lasso penalties are scale-dependent, so predictors are conventionally standardized before fitting to avoid arbitrarily favoring covariates measured in smaller units; \citet{HastieTibshiraniFriedman2009} and \citet{Bishop2006} present this as standard practice in applied statistical learning more broadly. In classical multivariate statistics, \citet{Jolliffe2002} documents the long-standing distinction between covariance-based and correlation-based (standardized) principal component analysis, motivated by exactly the same scale-sensitivity concern. \citet{KutnerNachtsheimNeterLi2005} and \citet{BelsleyKuhWelsch1980} further link covariate scale and centering to multicollinearity diagnostics and numerical conditioning of the design matrix in ordinary regression. Within survival analysis specifically, \citet{TherneauGrambsch2000} recommend centering covariates in the Cox partial likelihood to reduce numerical instability and improve the interpretability of the baseline hazard.

Taken together, this literature establishes a broad, qualitative consensus: covariate centering and scaling generally improve numerical conditioning and are recommended practice across linear, penalized, multivariate, and hazard-model settings. However, this consensus is prescriptive rather than empirical---it recommends \emph{that} covariates be scaled without quantifying \emph{which} scaling family should be preferred for a given model class, nor whether an accuracy-based comparison alone is sufficient to detect the resulting probabilistic quality of the fitted model. None of the cited works evaluate multiple named scaling families side by side under a formal convergence criterion and a multi-objective selection rule for a Markov deterioration hazard model. Standard practice recommends standardization for exactly the ill-conditioning reasons above, but the choice is rarely validated empirically against domain-specific alternatives such as range-based scaling or power transformations, and rarely audited for probability calibration once selected. The Box--Cox transformation~\citep{BoxCox1964} was originally proposed to induce approximate normality and variance stabilization in a response variable; its use here as a covariate preprocessing step, combined with a training-derived positivity shift to accommodate covariates that are not strictly positive, is a domain-specific adaptation whose optimization behavior we characterize empirically. This paper's contribution is precisely to close this gap for the Exponential Hazard Markov model: we quantify convergence rate, accuracy, optimizer time, and log-loss calibration across four scaling families on ten real member-case datasets, turning the general qualitative guidance summarized above into a specific, evidence-based decision rule (Section~\ref{sec:recommendation}).

\subsection{Probability Calibration and Proper Scoring Rules}

Accuracy (a 0--1 loss on the arg-max prediction) is only one facet of predictive quality. The Brier score~\citep{Brier1950} was among the first formal measures rewarding well-calibrated probabilistic forecasts rather than only correct arg-max predictions, and \citet{GneitingRaftery2007} generalize this into the theory of strictly proper scoring rules, including the logarithmic score (log loss), showing that optimizing or reporting accuracy alone can obscure severe miscalibration. Within machine learning specifically, \citet{Platt1999} and \citet{NiculescuMizilCaruana2005} document that classifiers achieving strong accuracy can nonetheless produce poorly calibrated probability estimates, motivating dedicated calibration diagnostics and post-hoc correction methods rather than relying on accuracy alone to certify predictive quality. This distinction is central to our findings: we observe scaling methods that tie on accuracy while differing by an order of magnitude in log loss, underscoring the need to report calibration-sensitive metrics alongside accuracy in any model-selection procedure, including the multi-objective, Pareto-based selection we adopt here.

\subsection{Multi-Objective Model Selection}

Model and hyperparameter selection under multiple, potentially conflicting objectives (e.g., accuracy versus computational cost) is naturally framed as a Pareto-optimality problem~\citep{Miettinen1999} rather than forcing a single scalarized objective, which would require an arbitrary trade-off weight chosen a priori. This framing is well established in multi-objective optimization more broadly, where algorithms such as NSGA-II~\citep{DebPratapAgarwalMeyarivan2002} are designed explicitly to recover the full non-dominated (Pareto) front rather than a single scalarized optimum, so that the trade-off between conflicting objectives is exposed to the decision-maker instead of being fixed in advance. We adopt this framing directly: rather than selecting a single ``best'' scaling method by a combined score, we report the full Pareto-non-dominated set of methods per member case with respect to golden-test accuracy and optimizer convergence time, deferring any further trade-off to the practitioner and explicitly separating this decision from the calibration guardrail discussed in Section~\ref{sec:discussion}.

\section{Methodology}
\label{sec:methodology}

This section presents the statistical model, the covariate scaling operator family, and the multi-objective selection rule in AI-STATS style: formal notation, definitions, an explicit algorithm, and a discussion of assumptions. Extended remarks, additional assumptions, and computational notes supporting this section are collected in Appendix~\ref{sec:appendix-methodology} for a self-contained presentation.

\subsection{Problem Setup and Notation}

Let $i = 1, \ldots, n$ index inspection transition records within a fixed member case. Each record consists of a first-inspection state $S_1^{(i)} \in \{1, \ldots, K\}$, a second-inspection state $S_2^{(i)} \in \{1, \ldots, K\}$ with $S_2^{(i)} \geq S_1^{(i)}$ (monotone, non-repairable degradation, enforced by upstream data cleaning), an inspection interval $\Delta t^{(i)} > 0$ in years, and a covariate vector $x^{(i)} \in \mathbb{R}^p$. The number of states $K$ is inferred empirically as the maximum observed rank in the training partition of each case. In this study, $p = 4$ with covariates \texttt{age\_at\_inspection\_1}, \texttt{bridge\_length}, \texttt{bridge\_area}, and \texttt{traffic\_mean}.

\subsection{Exponential Hazard Markov (EHM) Model}
\label{sec:ehm}

\textbf{Definition 1 (State-wise hazard rate).} For state $k \in \{1, \ldots, K-1\}$ and covariate vector $x$, the instantaneous hazard of leaving state $k$ is
\begin{equation}
\theta_k(x) = \exp\!\big(\tilde{x}^\top \beta_k\big), \qquad \tilde{x} = (1, x_1, \ldots, x_p)^\top,
\label{eq:hazard}
\end{equation}
where $\beta_k \in \mathbb{R}^{p+1}$ includes an intercept. State $K$ is absorbing.

\textbf{Definition 2 (Monotone birth-type generator).} The process is a continuous-time Markov chain restricted to adjacent forward transitions, with generator $Q_{kk}(x) = -\theta_k(x)$, $Q_{k,k+1}(x) = \theta_k(x)$ for $k < K$, and no transitions leaving state $K$.

\textbf{Proposition 1 (Closed-form transition probabilities).} For $i \leq j < K$ with distinct hazards, the transition probability over interval $\Delta t$ is
\begin{equation}
\pi_{ij}(\Delta t \mid x) = \sum_{k=i}^{j} \left(\prod_{\substack{m=i \\ m \neq k}}^{j} \frac{\theta_m(x)}{\theta_m(x) - \theta_k(x)}\right) \exp\!\big(-\theta_k(x)\, \Delta t\big),
\label{eq:transition-prob}
\end{equation}
with $\pi_{ii}(\Delta t \mid x) = \exp(-\theta_i(x)\Delta t)$ and $\pi_{iK}(\Delta t \mid x) = 1 - \sum_{j=i}^{K-1}\pi_{ij}(\Delta t \mid x)$ by complementarity. This is the standard closed form for a pure-birth (Coxian-type) continuous-time Markov chain with distinct rates. Near-coincident hazards are handled by a numerical regularization ($10^{-9}$ denominator floor) rather than a distinct closed form.

\textbf{Definition 3 (Likelihood and estimator).} Let $\beta = (\beta_1, \ldots, \beta_{K-1})$. The unweighted maximum likelihood estimator solves
\begin{equation}
\hat{\beta} = \arg\min_{\beta} \; -\sum_{i=1}^{n} \log \max\!\big(\pi_{S_1^{(i)} S_2^{(i)}}(\Delta t^{(i)} \mid x^{(i)}),\, 10^{-12}\big),
\label{eq:mle}
\end{equation}
via L-BFGS-B~\citep{ByrdLuNocedalZhu1995} with at most 500 iterations, initialized at $\beta^{(0)} \sim \mathcal{N}(0, 0.1^2 I)$.

\subsection{Covariate Scaling Operator Family}
\label{sec:scaling-family}

Let $x$ denote a single covariate column restricted to the training partition, with $x_{\min}$, $x_{\max}$, mean $\bar{x}$, and standard deviation $s$ (population, ddof$=0$). A scaling method $m$ defines a training-fitted operator $T_m(\cdot; \hat{\theta}_m)$ applied unchanged to both training and golden-test rows, preventing evaluation-partition leakage into the transformation parameters themselves.

\begin{align}
T_{\text{max}}(x) &= x / x_{\max} \label{eq:max} \\
T_{\text{minmax}}(x) &= (x - x_{\min}) / (x_{\max} - x_{\min}) \label{eq:minmax} \\
T_{\text{zscore}}(x) &= (x - \bar{x}) / s \label{eq:zscore} \\
T_{\text{boxcox}}(x) &= \operatorname{BoxCox}(x + s_{\text{shift}};\, \hat{\lambda}), \quad s_{\text{shift}} = \max(0,\, 1 - x_{\min}) \label{eq:boxcox}
\end{align}

where $\hat{\lambda}$ is estimated by profile maximum likelihood~\citep{BoxCox1964} on the shifted training column, and $\operatorname{BoxCox}(u; \lambda) = (u^{\lambda}-1)/\lambda$ for $\lambda \neq 0$, $\log u$ for $\lambda = 0$. Degenerate (constant) training columns are mapped to zero under all four methods, avoiding division by zero.

\subsection{Convergence Event and Time Measurement}
\label{sec:convergence}

A trial is said to \textbf{converge} if the L-BFGS-B result reports \texttt{success = True} and the returned parameters and log-likelihood are finite; otherwise it is recorded as \textbf{failed} with the solver's termination message, and excluded from Pareto selection (Section~\ref{sec:selection}) but retained in the reported comparison. Convergence time $\tau$ is a monotonic wall-clock measurement taken immediately before and after the single call to the optimizer, excluding dataset construction, scaling fit/transform, golden-test evaluation, and post-fit diagnostics (Hessian, AIC/BIC, plotting), so that $\tau$ isolates optimizer-conditioning effects attributable to scaling.

\subsection{Golden-Test Risk Functionals}

Let $\mathcal{G}_c$ denote golden-test rows for case $c$ with states supported by the training-inferred $K$ (unsupported rows are excluded and counted separately). Accuracy is the arg-max 0--1 loss,
\begin{equation}
\mathrm{Acc}(c, m) = \frac{1}{|\mathcal{G}_c|} \sum_{i \in \mathcal{G}_c} \mathbb{1}\!\left[ S_2^{(i)} = \arg\max_{j} \hat\pi_{S_1^{(i)} j}\big(\Delta t^{(i)} \mid T_m(x^{(i)})\big) \right],
\label{eq:accuracy}
\end{equation}
and log loss is the strictly proper score~\citep{GneitingRaftery2007}
\begin{equation}
\mathrm{LogLoss}(c, m) = -\frac{1}{|\mathcal{G}_c|} \sum_{i \in \mathcal{G}_c} \log \max\!\big(\hat\pi_{S_1^{(i)} S_2^{(i)}}\big(\Delta t^{(i)} \mid T_m(x^{(i)})\big),\, 10^{-12}\big).
\label{eq:logloss}
\end{equation}
Macro recall (mean per-transition-class accuracy) is also reported as a diagnostic. Only accuracy and $\tau$ enter the formal selection rule below; log loss and macro recall are used solely to characterize calibration and class-level behavior post hoc.

\subsection{Multi-Objective Selection via Pareto Dominance}
\label{sec:selection}

Let $\mathcal{E}_c$ be the set of methods that converge for case $c$, and define the objective vector $v(c, m) = (\mathrm{Acc}(c, m), -\tau(c, m))$ (both components maximized).

\textbf{Definition 4 (Pareto dominance).} $m \succ_c m'$ if $v(c, m) \geq v(c, m')$ component-wise with at least one strict inequality.

\textbf{Definition 5 (Selected set).}
\begin{equation}
\mathrm{Sel}(c) = \{ m \in \mathcal{E}_c : \nexists\, m' \in \mathcal{E}_c \text{ with } m' \succ_c m \}.
\label{eq:pareto-set}
\end{equation}
All non-dominated methods are retained per case; the rule does not force a single winner. Floating-point comparisons use absolute tolerances of $10^{-12}$ (accuracy) and $10^{-9}$ seconds (duration).

\begin{algorithm}[H]
\caption{Scaling comparison and Pareto selection}
\label{alg:comparison}
\begin{algorithmic}[1]
\STATE \textbf{Input}: raw inspection tables, member-case definitions, seed, row cap
\STATE \textbf{Output}: per-trial comparison table, per-case selected-method set
\STATE Build ten member-case datasets once, shared across all methods
\FOR{each member case $c$}
    \STATE Split case data into training and golden-test partitions
    \FOR{each method $m \in \{\text{max}, \text{minmax}, \text{zscore}, \text{boxcox}\}$}
        \STATE Fit $T_m$ on the training partition only (Eqs.~\ref{eq:max}--\ref{eq:boxcox})
        \STATE Transform training covariates; reset the estimation random seed
        \STATE Fit the EHM model via L-BFGS-B (Eq.~\ref{eq:mle}); record convergence and $\tau$
        \IF{converged with finite parameters and log-likelihood}
            \STATE Transform golden-test covariates with the \emph{same} fitted $T_m$
            \STATE Compute $\mathrm{Acc}(c,m)$, $\mathrm{LogLoss}(c,m)$, macro recall (Eqs.~\ref{eq:accuracy}--\ref{eq:logloss})
        \ELSE
            \STATE Record failure status and solver message; mark metrics as missing
        \ENDIF
    \ENDFOR
    \STATE Compute $\mathrm{Sel}(c)$ over converged methods (Eq.~\ref{eq:pareto-set})
\ENDFOR
\STATE Aggregate all per-case, per-method rows into the comparison table
\end{algorithmic}
\end{algorithm}

\subsection{Assumptions and Threats}

The generator in Definition 2 assumes genuinely monotone degradation and would be misspecified under repair or measurement-driven downward transitions. Scaling parameters are fit exclusively on the training partition (preventing leakage into $T_m$), but the golden test is nonetheless used for method \emph{selection} in this study; consequently, reported post-selection accuracy is optimistic relative to a fully independent hold-out evaluation, a limitation we return to in Section~\ref{sec:discussion}. The Box-Cox positivity shift is fit on training data only, so golden-test covariates falling outside the training range are not guaranteed positive under the shift, which the implementation surfaces as an explicit trial failure rather than silent clipping.

\section{Experimental Setup}
\label{sec:experiments}

\subsection{Dataset Description}

We analyze anonymized bridge inspection transition data derived from a bridge inventory of paired inspection records (DATLINKSET-I/II/III), covering ten member/damage categories used in prior production modeling: Main girder, Pier, Bearing, Deck slab, Cross girder, Longitudinal girder, Abutment, Expansion joint, Slope protection, and Pavement. For each case, third-to-fourth inspection transitions form the training partition; fourth-to-fifth inspection transitions are split at the bridge level (no bridge appears in both partitions) into a 20\% validation partition (not used in this study) and an 80\% golden-test partition, using a fixed random seed of $20260924$. A row cap of 20,000 is applied per case, matching the production ingestion limit.

Table~\ref{tab:cases} summarizes the ten member cases under these conditions, which exactly replicate the prior unweighted production baseline.

\begin{table}[h]
\centering
\caption{Member-case sample sizes under the replicated baseline conditions (seed $20260924$, 4 covariates, unweighted likelihood).}
\label{tab:cases}
\begin{tabular}{lrrrr}
\toprule
\textbf{Member case} & \textbf{Total rows} & \textbf{Train} & \textbf{Validation} & \textbf{Golden test} \\
\midrule
Main girder & 2,409 & 624 & 217 & 1,568 \\
Pier & 3,087 & 1,158 & 285 & 1,644 \\
Bearing & 4,269 & 1,634 & 640 & 1,995 \\
Deck slab & 3,478 & 1,491 & 325 & 1,662 \\
Cross girder & 604 & 216 & 113 & 275 \\
Longitudinal girder & 272 & 86 & 41 & 145 \\
Abutment & 800 & 271 & 91 & 438 \\
Expansion joint & 779 & 323 & 105 & 351 \\
Slope protection & 79 & 27 & 14 & 38 \\
Pavement & 1,924 & 742 & 208 & 974 \\
\bottomrule
\end{tabular}
\end{table}

\subsection{Explanatory Covariates}

All ten cases use the same four explanatory covariates: \texttt{age\_at\_inspection\_1} (years since completion at the first inspection), \texttt{bridge\_length} (meters), \texttt{bridge\_area} (square meters), and \texttt{traffic\_mean} (average of the third- and fourth-inspection traffic volumes, vehicles per day). These covariates differ by several orders of magnitude in native units, motivating the scaling comparison in this study.

\subsection{Experimental Conditions}

All 40 trials (10 cases $\times$ 4 methods) share the following fixed conditions, replicating the prior production baseline exactly except for the manipulated scaling method: unweighted likelihood (no class-frequency reweighting), the same bridge-level train/golden-test split and seed, the same row cap, and the same four covariates. Each trial resets the estimation random seed immediately before model fitting, so that differences in optimizer trajectories across methods are attributable to the scaled covariate values rather than differing initializations.

\subsection{Implementation and Software}

\begin{itemize}
    \item \textbf{Programming language}: Python 3.13
    \item \textbf{Optimization}: SciPy 1.16.3~\citep{Virtanen2020} for \texttt{scipy.optimize.minimize} (L-BFGS-B, \texttt{maxiter=500}) and \texttt{scipy.stats.boxcox}
    \item \textbf{Model implementation}: the production \texttt{ExponentialHazardMarkovModel} class, extended in this study to record optimizer success, termination message, iteration count, function-evaluation count, and wall-clock duration
    \item \textbf{Scaling, evaluation, and selection}: a dedicated internal module implementing the four scaling operators (Section~\ref{sec:scaling-family}), golden-test evaluation (Section~\ref{sec:ehm}), and the Pareto selection procedure (Section~\ref{sec:selection})
    \item \textbf{Testing}: an automated unit-test suite covering scaler correctness on synthetic data (including a constant column and a column requiring a Box-Cox positivity shift), metadata round-tripping, and Pareto selection under ties and complete failure
\end{itemize}

\subsection{Reproducibility}

All results in this paper derive from a single deterministic run with seed $20260924$ and the row cap described above; the same procedure reproduces the full 40-trial comparison table and per-case selection summaries used throughout Section~\ref{sec:results}. Convergence time is a wall-clock measurement and is therefore hardware-dependent; the relative ranking of methods, not the absolute timing, is the object of interest in this study. Code and the anonymized dataset are retained internally for confidentiality and are not distributed externally as part of this paper.

\section{Results}
\label{sec:results}

\subsection{Convergence, Accuracy, and Timing by Method}

Table~\ref{tab:method-summary} summarizes, for each scaling method across all ten member cases, the number of converged trials (Section~\ref{sec:convergence}), the number of Pareto-selected cases (Section~\ref{sec:selection}), the mean golden-test accuracy among converged trials, and the median optimizer-only convergence time.

\begin{table}[h]
\centering
\caption{Method-level summary across 10 member cases (40 trials total; 32 converged, 11 case-level Pareto selections).}
\label{tab:method-summary}
\begin{tabular}{lrrrr}
\toprule
\textbf{Method} & \textbf{Converged} & \textbf{Selected} & \textbf{Mean accuracy} & \textbf{Median time (s)} \\
\midrule
Max scaling (baseline) & 9 / 10 & 0 & 0.5174 & 0.928 \\
Min-max scaling & 9 / 10 & 1 & 0.5174 & 0.988 \\
Z-score scaling & 10 / 10 & 7 & 0.5255 & 0.358 \\
Box-Cox transformation & 4 / 10 & 3 & 0.4546 & 0.110 \\
\bottomrule
\end{tabular}
\end{table}

Z-score scaling is the only method that converges for all ten member cases. Among the three methods that converge for the majority of cases (max, min-max, z-score), mean converged accuracy differs by at most 0.008, while median convergence time for z-score is less than 40\% of max or min-max scaling. Box-Cox converges in only 4 of 10 cases; its median time among converged trials is the lowest of all four methods, but this reflects a small, non-representative subset of cases (Section~\ref{sec:results-boxcox}).

\begin{figure}[ht]
\centering
\includegraphics[width=0.95\textwidth]{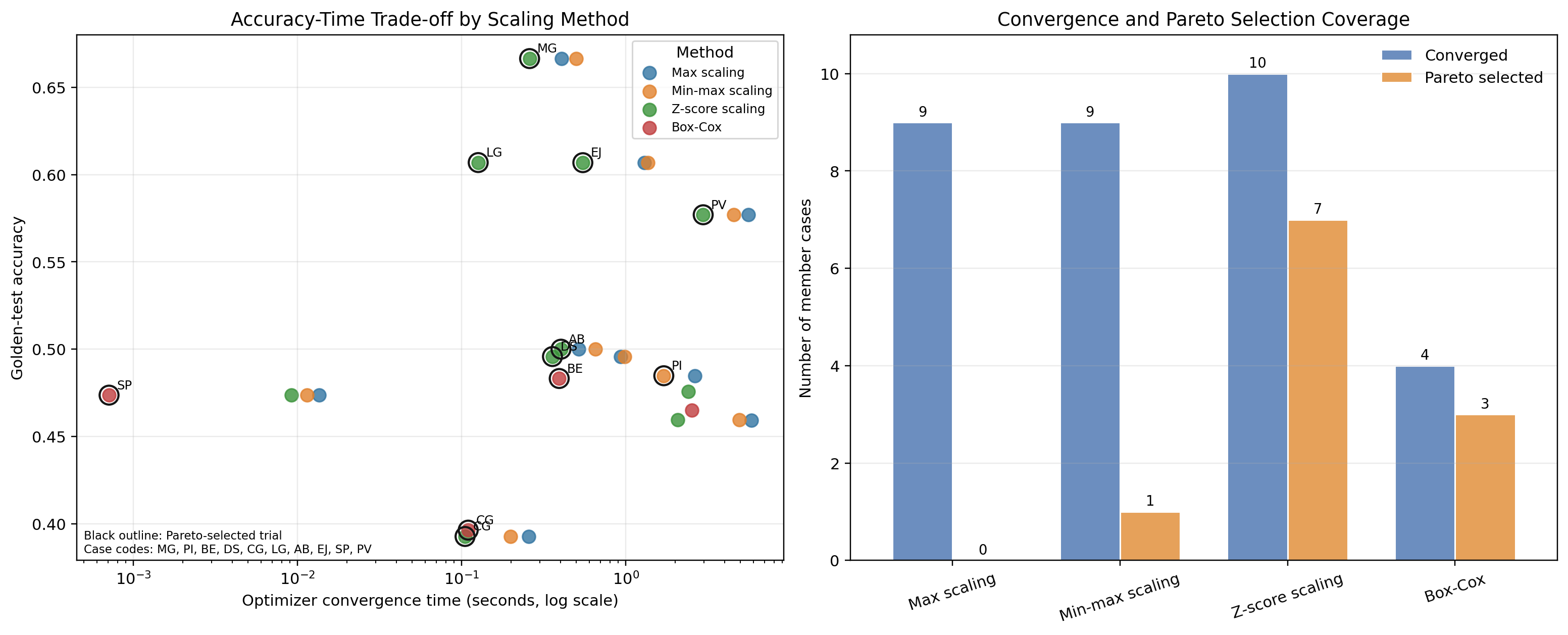}
\caption{Accuracy--time trade-off (left) and convergence/selection coverage (right) across all 40 trials. Black-outlined markers indicate Pareto-selected trials. Z-score scaling (green) converges for every case and occupies the low-time, high-accuracy region for most cases.}
\label{fig:tradeoff}
\end{figure}

\subsection{Per-Case Pareto Selection}

Table~\ref{tab:selection} lists the Pareto-selected method(s) for each of the ten member cases. Z-score scaling is selected in 7 of 10 cases; Box-Cox is selected in 3 cases (Bearing, Cross girder, Slope protection); min-max is selected in 1 case (Pier); max scaling is never selected on its own. Cross girder is the only case with two co-selected, mutually non-dominated methods (z-score and Box-Cox), since Box-Cox is both slightly more accurate and no slower in that specific case.

\begin{table}[h]
\centering
\caption{Pareto-selected scaling method(s) per member case.}
\label{tab:selection}
\begin{tabular}{ll}
\toprule
\textbf{Member case} & \textbf{Selected method(s)} \\
\midrule
Main girder & z-score \\
Pier & min-max \\
Bearing & Box-Cox \\
Deck slab & z-score \\
Cross girder & z-score, Box-Cox \\
Longitudinal girder & z-score \\
Abutment & z-score \\
Expansion joint & z-score \\
Slope protection & Box-Cox \\
Pavement & z-score \\
\bottomrule
\end{tabular}
\end{table}

\begin{figure}[ht]
\centering
\includegraphics[width=0.95\textwidth]{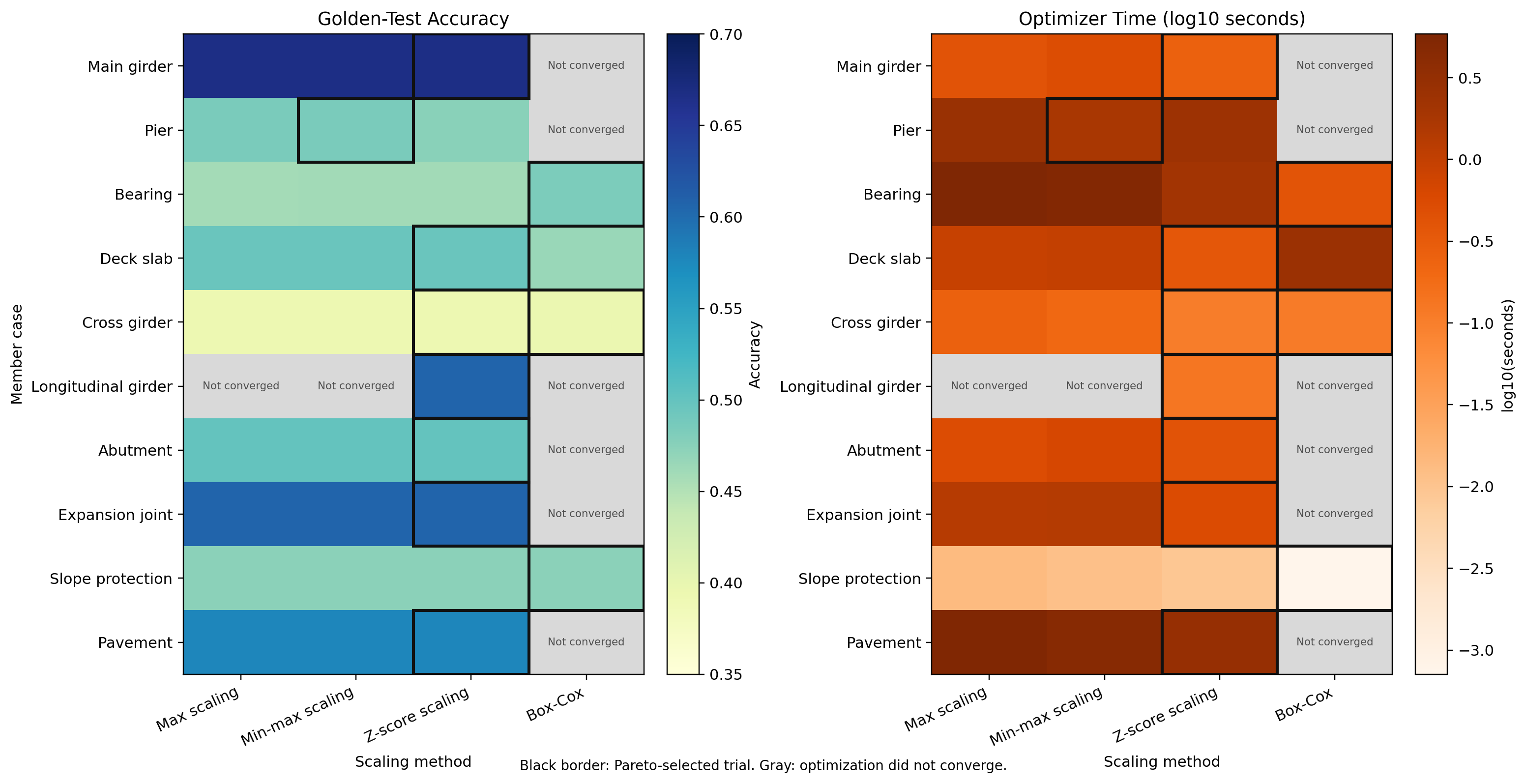}
\caption{Per-case, per-method golden-test accuracy (left) and $\log_{10}$ optimizer time in seconds (right). Gray cells indicate non-convergence; black-outlined cells indicate the Pareto-selected trial(s) for that case.}
\label{fig:heatmaps}
\end{figure}

Longitudinal girder (86 training rows) and Slope protection (27 training rows) are the two smallest cases; both show at least two of four methods failing to converge (max and min-max for Longitudinal girder; max, min-max, and z-score all converge for Slope protection but Box-Cox alone is selected there by strictly lower time at tied accuracy).

\subsection{Box-Cox Convergence and Calibration}
\label{sec:results-boxcox}

Box-Cox scaling converges in only 4 of 10 cases (Bearing, Cross girder, Slope protection, and one additional non-selected case), a markedly lower convergence rate than the other three methods. Table~\ref{tab:boxcox-logloss} compares golden-test log loss for the three cases where Box-Cox is Pareto-selected against the log-loss range of the alternative methods on the same case.

\begin{table}[h]
\centering
\caption{Box-Cox golden-test accuracy and log loss versus alternative methods, for the three cases where Box-Cox is Pareto-selected.}
\label{tab:boxcox-logloss}
\begin{tabular}{lrrr}
\toprule
\textbf{Member case} & \textbf{Box-Cox accuracy} & \textbf{Box-Cox log loss} & \textbf{Alternatives' log loss} \\
\midrule
Bearing & 0.4832 & 14.2609 & 1.9758--2.0454 \\
Cross girder & 0.3964 & 16.6358 & 2.9707--3.4432 \\
Slope protection & 0.4737 & 13.8499 & $\approx$1.2810 \\
\bottomrule
\end{tabular}
\end{table}

In all three cases, Box-Cox accuracy is within 0.03 of the alternatives (and non-dominated under the Pareto rule), yet its log loss is 7--13$\times$ worse. This demonstrates directly that the two-objective (accuracy, time) selection rule alone can select a severely miscalibrated model whenever it is not strictly dominated on the two chosen objectives; log loss is required to detect this failure mode, but is not, by design, part of the formal selection criterion in this study (Section~\ref{sec:selection}).

\subsection{Summary of Aggregate Outcomes}

Of the 40 trials, 32 (80\%) converge and 11 case-level Pareto selections are recorded across the 10 cases (one case yields two co-selected methods). Z-score scaling accounts for 7 of these 11 selections and is the only method with zero convergence failures, making it the most broadly reliable candidate under the conditions replicated in this study.

\section{Discussion}
\label{sec:discussion}

\subsection{Why Z-Score Scaling Outperforms Range-Based Scaling}

Max scaling and min-max scaling both rescale a covariate's range without centering it, so that covariates with very different native units (component age in tens of years versus traffic volume in tens of thousands of vehicles per day) remain on substantially different effective scales relative to their variance. This is precisely the condition under which quasi-Newton optimizers such as L-BFGS-B~\citep{ByrdLuNocedalZhu1995} experience poor Hessian conditioning and slower or failed convergence~\citep{NocedalWright2006}. Z-score scaling both centers and equalizes the variance of each covariate, which directly addresses this conditioning problem; the empirical result---z-score converging in all 10 cases with less than half the median optimizer time of max or min-max scaling---is consistent with this mechanism rather than being merely coincidental.

\subsection{Why Box-Cox Scaling Is Fast but Fragile}

Box-Cox scaling applies a nonlinear, training-fitted power transformation after a positivity shift (Eq.~\ref{eq:boxcox}), which can substantially compress or expand the effective range of a covariate depending on the estimated shape parameter $\hat{\lambda}$. When it converges, the resulting covariate scale can be highly favorable to the optimizer, explaining the very low convergence times observed (Table~\ref{tab:method-summary}). However, this same nonlinear reshaping is estimated from a small training sample per case (as few as 27 rows for Slope protection) and is then applied unchanged to golden-test covariates that may fall outside the training range, both a source of numerical fragility (explaining the 4/10 convergence rate) and a plausible mechanism for the severe log-loss deterioration observed in the cases where it does converge (Section~\ref{sec:results-boxcox}): a transformation well-suited to compressing training-set variation is not guaranteed to preserve the covariate-to-hazard relationship needed for well-calibrated golden-test probabilities.

\subsection{Accuracy and Convergence Time Are an Incomplete Selection Basis}

The central empirical warning of this study is that a two-objective Pareto rule over accuracy and convergence time---a natural, defensible formalization of efficacy and efficiency---can nonetheless select a severely miscalibrated model whenever no other converged method strictly dominates it. Box-Cox's selection in three cases, despite log loss 7--13$\times$ worse than the alternatives, illustrates this precisely: accuracy (a 0--1 loss) is blind to how confidently a model favors the wrong or right prediction, whereas log loss (a strictly proper scoring rule~\citep{GneitingRaftery2007}) is not. We therefore do not recommend deploying the formal Pareto selection rule of Section~\ref{sec:selection} unguarded; Section~\ref{sec:recommendation} gives a practical amendment.

\subsection{Small-Sample Fragility}

The two smallest member cases (Slope protection, 27 training rows; Longitudinal girder, 86 training rows) both show elevated convergence failure rates relative to larger cases, consistent with Hypothesis H5 of the pre-specified experimental design (Appendix~\ref{sec:appendix-experiment-plan}). For these cases, any single-run comparison (including this one) should be treated as exploratory; repeated resampling or cross-validation would be needed to obtain stable estimates of both convergence probability and golden-test accuracy.

\subsection{Practical Recommendation}
\label{sec:recommendation}

Based on the combined evidence of reliability (convergence rate), efficacy (accuracy), efficiency (convergence time), and calibration (log loss), we recommend the following staged decision rule for this model class:

\begin{enumerate}
    \item Exclude non-converged or non-finite trials outright.
    \item Adopt z-score scaling as the default candidate for every member case, given its universal convergence and highest mean accuracy among converged trials.
    \item Retain an alternative scaling method only if its accuracy is not lower, its log loss is not materially worse, and its convergence is stable (not a single-run artifact), reserving the accuracy/time Pareto comparison for the final tie-break among methods that already pass the calibration guardrail.
    \item Confirm the chosen method once on an untouched, independent test partition, since golden-test-based selection in this study is optimistic (Section~\ref{sec:limitations}).
\end{enumerate}

\subsection{Limitations}
\label{sec:limitations}

\textbf{Selection-on-test.} The golden test is used directly for method selection in this study, by explicit design (Appendix~\ref{sec:appendix-experiment-plan}). Reported post-selection accuracy is therefore an optimistic estimate of generalization performance, not an independent confirmation; a genuinely held-out test partition, untouched during selection, would be required for a final performance claim.

\textbf{Single-seed timing.} Convergence time is measured under a single random seed and a single hardware configuration; absolute timings are hardware-dependent, though we expect the qualitative ranking (z-score fastest among reliably converging methods) to be robust to seed and hardware changes, since it follows from the conditioning argument in Section~\ref{sec:discussion} rather than from a particular random draw.

\textbf{Two-objective selection criterion.} By design, the formal Pareto rule (Section~\ref{sec:selection}) uses only accuracy and convergence time; it does not include log loss or macro recall as selection objectives, which is precisely why it can select a miscalibrated method. This is intentional---to demonstrate the gap empirically---rather than a proposed final production rule; Section~\ref{sec:recommendation} gives the amended rule we recommend for deployment.

\section{Conclusion}
\label{sec:conclusion}

We formalized covariate scaling as a training-fitted operator family for a Markov deterioration hazard model and compared four methods---baseline max scaling, min-max scaling, z-score scaling, and Box-Cox transformation---under conditions that exactly replicate an existing production baseline: ten bridge member/damage categories, four explanatory covariates, unweighted likelihood, and a fixed bridge-level train/golden-test split. Using a tolerance-controlled, two-objective Pareto selection rule over golden-test accuracy and optimizer convergence time, we found that z-score scaling converges in all ten member cases, achieves the highest mean converged accuracy, and requires less than half the median optimizer time of the current production baseline, making it Pareto-selected in 7 of 10 cases.

At the same time, our results demonstrate a specific and important failure mode of accuracy-and-time-only selection: Box-Cox scaling converges in only 4 of 10 cases, but where it does converge it is often extremely fast and Pareto-selected, while its golden-test log loss is 7--13$\times$ worse than the alternatives despite comparable accuracy. This gap between a 0--1 loss (accuracy) and a strictly proper scoring rule (log loss) is invisible to the formal Pareto rule as specified, and would not have been detected without explicitly reporting calibration-sensitive diagnostics alongside the two selection objectives.

We therefore recommend, for this model class, adopting z-score scaling as the default candidate, adding a log-loss non-deterioration guardrail before accepting any faster alternative, and confirming the final chosen method on an independent test partition rather than the golden test used for selection in this study. More broadly, this work illustrates a general methodological point for AI-STATS practice: multi-objective model-selection rules should be audited against metrics outside the selection objective itself, since Pareto-optimality with respect to a chosen pair of objectives provides no guarantee of quality along dimensions---such as probabilistic calibration---that were not included in the rule.

Future work includes validating scaling-method stability across repeated resampling seeds (particularly for the two smallest member cases), extending the formal selection rule to incorporate a log-loss guardrail directly rather than as a post hoc check, and evaluating whether the same conditioning argument (centering plus variance equalization) generalizes to other Markov hazard model families beyond the exponential-hazard specification studied here.

\bibliographystyle{unsrtnat}
\bibliography{references}

\clearpage
\appendix
\section{Additional Details}

\subsection{Extended Methodology Notes}
\label{sec:appendix-methodology}

This subsection collects remarks, additional assumptions, and computational notes that support Section~\ref{sec:methodology}, presented here in full so that the paper remains self-contained.

\textbf{Remark A.1 (Relation to proportional hazards).} The Exponential Hazard Markov model of Section~\ref{sec:ehm} is a discretized, multi-state extension of the proportional-hazards idea (a state-specific log-linear hazard in the covariates), specialized to an ordered, absorbing, non-repairable state space appropriate for infrastructure deterioration, rather than a general multi-state model permitting recovery or competing risks.

\textbf{Remark A.2 (Class weighting is out of scope).} A class-weighted variant of the likelihood in Eq.~\eqref{eq:mle} exists in the production tool (inverse transition-frequency weights) but is intentionally not used in this study; all 40 trials are fit with uniform weights, matching the unweighted production baseline described in Section~\ref{sec:experiments}.

\textbf{Remark A.3 (Accuracy versus log loss).} Accuracy (Eq.~\eqref{eq:accuracy}) evaluates only the arg-max prediction (a 0--1 loss), whereas log loss (Eq.~\eqref{eq:logloss}) is a strictly proper scoring rule sensitive to the entire predicted probability vector~\citep{GneitingRaftery2007}. A method can tie or nearly tie on accuracy while being severely miscalibrated in log loss (Section~\ref{sec:results-boxcox}); this distinction motivates reporting both, even though only accuracy and convergence time enter the formal selection rule (Section~\ref{sec:selection}).

\textbf{Assumption A.4 (Monotone degradation).} $S_2 \geq S_1$ is enforced by upstream data cleaning; the generator of Section~\ref{sec:ehm} would be misspecified for data containing genuine repairs or measurement-driven downward transitions.

\textbf{Assumption A.5 (Train-only scaling fit and selection-on-test).} Scaling parameters are estimated exclusively from the training partition (Section~\ref{sec:scaling-family}), preventing leakage into $T_m$ itself; however, the golden test is still used for method \emph{selection} (Section~\ref{sec:selection}), so the reported post-selection accuracy is optimistic relative to a fully held-out evaluation (Section~\ref{sec:limitations}).

\textbf{Assumption A.6 (Box-Cox extrapolation).} The positivity shift $s_{\text{shift}}$ (Eq.~\eqref{eq:boxcox}) is fit on training data only; if golden-test covariates fall outside the training range, $x + s_{\text{shift}}$ is not guaranteed positive, which the implementation treats as an explicit trial failure rather than silently clipping.

\textbf{Assumption A.7 (Distinct hazards).} Proposition 1 (Section~\ref{sec:ehm}) assumes distinct $\theta_k(x)$ values; near-coincident hazards are handled by a numerical floor ($10^{-9}$ denominator) rather than a distinct closed form, which can reduce numerical precision without causing failure.

\textbf{Computational notes.} All trials run under Python 3.13 with SciPy 1.16.3~\citep{Virtanen2020} for \path{scipy.optimize.minimize} (L-BFGS-B) and \path{scipy.stats.boxcox}. The model class and optimizer diagnostics, and the scaling, evaluation, and selection procedures, are implemented in an internal production codebase retained for confidentiality (Section~\ref{sec:experiments}).

\subsection{Pre-Specified Experimental Design and Hypotheses}
\label{sec:appendix-experiment-plan}

This subsection presents, in full, the pre-specified experimental design and hypotheses that guided the implementation described in Section~\ref{sec:experiments}, written in pre-registration style for reproducibility. It is a retrospective formalization, not a blinded ex-ante registration.

\textbf{Hypotheses.}
\begin{itemize}
    \item \textbf{H1 (reliability)}: z-score scaling converges for all ten member cases, while max, min-max, and Box-Cox scaling fail to converge in at least one case each.
    \item \textbf{H2 (accuracy)}: among methods that converge for a given case, differences in golden-test accuracy are small, so accuracy alone is an insufficient basis for method selection.
    \item \textbf{H3 (speed-calibration trade-off)}: Box-Cox scaling, when it converges, achieves the fastest optimizer convergence but exhibits materially worse log loss than max, min-max, or z-score scaling on the same case.
    \item \textbf{H4 (Pareto concentration)}: under the two-objective Pareto rule (maximize accuracy, minimize convergence time), z-score scaling is non-dominated in a majority of the ten member cases.
    \item \textbf{H5 (small-sample fragility)}: member cases with fewer than 100 training rows show a higher proportion of non-converged trials than cases with more than 500 training rows.
\end{itemize}

\textbf{Pre-specified decision rule.} A trial $(c, m)$ is \emph{eligible} if it converged with finite log-likelihood and parameters. Among eligible trials for case $c$, method $m$ is \emph{selected} if no other eligible method $m'$ satisfies accuracy$(c, m') \geq$ accuracy$(c, m)$ and time$(c, m') \leq$ time$(c, m)$ with at least one strict inequality (Eq.~\eqref{eq:pareto-set}); all non-dominated methods are retained and no single winner is forced.

\textbf{Threats to validity.} (1) \emph{Selection-on-test}: the golden test is used directly for method selection, so reported post-selection accuracy is optimistic, not an independent generalization estimate. (2) \emph{Small-sample instability}: cases with under 100 training rows (Slope protection: 27, Longitudinal girder: 86) yield noisy accuracy and log-loss estimates; conclusions for these cases are exploratory. (3) \emph{Single-seed design}: one random initialization seed is used per model fit; absolute convergence time may vary under different seeds or hardware, though the relative ranking across methods is the object of interest, not absolute timing. (4) \emph{Metric incompleteness}: accuracy and convergence time do not capture probability calibration by themselves; log loss is reported precisely to surface this gap (H3).

\textbf{Outcome summary (post hoc).} All five hypotheses were empirically supported: H1 (z-score converged 10/10; max and min-max converged 9/10 each, both failing on Longitudinal girder; Box-Cox converged only 4/10), H2 (among max, min-max, and z-score, accuracy differs by at most 0.01 within a case), H3 (Box-Cox log loss reaches 13.85--16.64 in three selected cases versus 1.28--3.44 for the alternatives, while accuracy differs by at most 0.03), H4 (z-score is selected in 7 of 10 cases), and H5 (both cases with fewer than 100 training rows have at least two of four methods fail to converge, versus at most one failure in every case with more than 500 training rows).

\subsection{Full Per-Trial Comparison Table}

Table~\ref{tab:appendix-full} reports golden-test accuracy and optimizer convergence time for every trial that converged, across all ten member cases and four scaling methods. Cells marked ``---'' indicate non-convergence (see Table~\ref{tab:method-summary} for aggregate convergence counts).

\begin{table}[h]
\centering
\footnotesize
\caption{Golden-test accuracy / optimizer time (seconds) for every trial. Bold indicates the Pareto-selected method(s) for that case.}
\label{tab:appendix-full}
\begin{tabular}{lrrrr}
\toprule
\textbf{Member case} & \textbf{Max} & \textbf{Min-max} & \textbf{Z-score} & \textbf{Box-Cox} \\
\midrule
Main girder & 0.6665 / 0.408 & 0.6665 / 0.498 & \textbf{0.6665 / 0.260} & --- \\
Pier & 0.4848 / 2.638 & \textbf{0.4848 / 1.705} & 0.4757 / 2.413 & --- \\
Bearing & 0.4591 / 5.842 & 0.4596 / 4.902 & 0.4596 / 2.079 & \textbf{0.4832 / 0.393} \\
Deck slab & 0.4958 / 0.928 & 0.4958 / 0.988 & \textbf{0.4958 / 0.358} & 0.4651 / 2.525 \\
Cross girder & 0.3927 / 0.257 & 0.3927 / 0.199 & \textbf{0.3927 / 0.105} & \textbf{0.3964 / 0.110} \\
Longitudinal girder & --- & --- & \textbf{0.6069 / 0.126} & --- \\
Abutment & 0.5000 / 0.517 & 0.5000 / 0.655 & \textbf{0.5000 / 0.403} & --- \\
Expansion joint & 0.6068 / 1.296 & 0.6068 / 1.363 & \textbf{0.6068 / 0.548} & --- \\
Slope protection & 0.4737 / 0.0135 & 0.4737 / 0.0114 & 0.4737 / 0.0092 & \textbf{0.4737 / 0.0007} \\
Pavement & 0.5770 / 5.591 & 0.5770 / 4.565 & \textbf{0.5770 / 2.963} & --- \\
\bottomrule
\end{tabular}
\end{table}

\subsection{Software Versions}

Python 3.13; SciPy 1.16.3; pandas 2.2.3; NumPy 2.3.5; numdifftools 0.9.42 (used for Hessian-based standard errors in the underlying model class, not for scaling-method selection).

\subsection{Relationship to Prior Production Baseline}

All experimental conditions in this study---member-case definitions, covariate set, unweighted likelihood, random seed, and row cap---replicate a prior internal production baseline that used max scaling exclusively for all ten member cases. This study extends that baseline by introducing three alternative scaling methods (min-max, z-score, Box-Cox) and the Pareto selection procedure described in Section~\ref{sec:methodology}; the ``Max'' column of Table~\ref{tab:appendix-full} reports this study's own max-scaling trials under the identical conditions, without modifying the prior baseline's own recorded outputs.

\end{document}